\documentclass[pdflatex,sn-mathphys,Numbered]{sn-jnl}
\usepackage{graphicx}%
\usepackage{multirow}%
\usepackage{amsmath,amssymb,amsfonts}%
\usepackage{amsthm}%
\usepackage{mathrsfs}%
\usepackage[title]{appendix}%
\usepackage{xcolor}%
\usepackage{textcomp}%
\usepackage{manyfoot}%
\usepackage{booktabs}%
\usepackage{algorithm}%
\usepackage{algorithmicx}%
\usepackage{algpseudocode}%
\usepackage{listings}%
\usepackage{geometry}
\DeclareMathOperator{\sech}{sech}

\usepackage{xr-hyper}

\begin{document}

\title[Article Title]{Electrically-injected room-temperature waveguide polariton laser}


\author[1]{\fnm{H.} \sur{Souissi}}

\author[2]{\fnm{V.} \sur{Develay}}

\author[3]{\fnm{M.} \sur{Gromovyi}}

\author[1]{\fnm{E.} \sur{Cambril}}

\author[2]{\fnm{C.} \sur{Brimont}}

\author[2]{\fnm{L.} \sur{Doyennette}}

\author[4]{\fnm{G.} \sur{Malpuech}}

\author[4,5]{\fnm{D.~D.} \sur{Solnyshkov}}

\author[3]{\fnm{B.} \sur{Alloing}}

\author[3]{\fnm{S.} \sur{Chenot}}

\author[3]{\fnm{M.} \sur{Al-Khalfioui}}

\author[3]{\fnm{E.} \sur{Frayssinet}}

\author[3]{\fnm{J.Y.} \sur{Duboz}}

\author[3]{\fnm{M.} \sur{Zhang}}

\author*[3,6]{\fnm{J.} \sur{Z\'u\~niga-P\'erez}}\email{jesus.zuniga-perez@cnrs.fr}

\author*[1]{\fnm{S.} \sur{Bouchoule}}\email{sophie.bouchoule@cnrs.fr}

\author*[2]{\fnm{T.} \sur{Guillet}}\email{thierry.guillet@umontpellier.fr}

\affil[1]{\orgdiv{Centre de Nanosciences et de Nanotechnologies}, \orgname{CNRS, Universit\'e Paris-Saclay}, \country{France}}

\affil[2]{\orgdiv{Laboratoire Charles Coulomb}, \orgname{Universit\'e de Montpellier, CNRS}, \city{Montpellier}, \country{France}}

\affil[3]{\orgdiv{University Côte d’Azur}, \orgname{CNRS, CRHEA}, \state{Valbonne}, \country{France}}

\affil[4]{\orgname{Institut Pascal, PHOTON-N2},\orgdiv{Universit\'e Clermont Auvergne, CNRS, Clermont INP}, \state{Clermont-Ferrand}, \country{France}}

\affil[5]{\orgdiv{Institut Universitaire de France (IUF)}, \state{Paris}, \country{France}}

\affil[6]{\orgdiv{MajuLab, International Research Laboratory IRL 3654}, \orgname{CNRS, Université Côte d’Azur}, \state{Sorbonne Université, National University of Singapore, Nanyang Technological University, Singapore}, \country{Singapore}}


\abstract{Exciton-polariton lasers are coherent light sources which do not require the population inversion (transparency) condition to be fulfilled. They have been conceptualized at the end of the XXth century but until now they operate almost exclusively under optical injection, which severely limits the widespread integration of the polariton-based devices implemented so far. Here we tackle this issue by reporting an electrically-pumped exciton-polariton laser based on GaN and operating at room temperature in a mode-locked regime. The laser architecture is close to the geometry of commercial ridge-waveguide GaN lasers, but based on a bulk GaN active region instead of quantum wells. Unique features of polariton lasers are demonstrated, in particular the breakdown of the transparency condition, which enables our polariton lasers to operate even when only a small fraction (20\%) of the cavity length is injected. Moreover, the large polaritonic gain allows for the operation of a short cavity length (60~µm) compared  to  commercial  lasers. From the very same sample, we also achieve polariton lasing under optical injection, confirming that the doped layers necessary for electrical injection do not prevent strong-coupling nor polariton lasing. Our results open a new perspective for polariton-based devices.
}

\maketitle

\section*{Introduction}\label{sec1}

To cope with the stringent requirements for on-chip optical signal processing, the number of functionalities within photonic integrated  circuits (PICs) keeps increasing  continuously \cite{Kitayama_2019}. Passive devices, based on materials that are transparent at the operation wavelength \cite{Stern_2015}, provide frequency selection and spectral shaping thanks to resonators, waveguides and beam splitters. Advanced integrated light sources, including lasers \cite{guang-hua_duan_hybrid_2014, Xiang_2021} and amplifiers \cite{Davenport_2016}, require active layers that provide enough gain and strong optical nonlinearities, enabling frequency-conversion \cite{Wang_2018} and complete spectral and temporal control of the laser emission. This has led to the development of optical frequency combs \cite{Del_Haye_2007}, super-continuum \cite{Zhao_2015} and dual-frequency integrated optical sources. Relying on the combination of active and passive building blocks, PICs are presently extending to wardsmore complex architectures. These systems are able to address an increasing number of applications, from sensing to quantum and neuromorphic computing, to all-optical classical and quantum information processing \cite{Bogaerts_2020}. Such dense integration of photonic devices requires elements with smaller spatial footprint and reduced thermal budget, which, in turn, imposes active layers with even larger gain and stronger nonlinearities. Ultimately, nonlinear optical systems might attain the regime where small photon numbers are sufficient to establish the necessary nonlinear response \cite{Benimetskiy2025}.

Exciton-polaritons, hybrid light-matter quasiparticles arising from the strong coupling between semiconductor excitons and confined light in resonators \cite{weisbuch_observation_1992,kavokin_microcavities_2006}, are extremely appealing in this context due to their intrinsic strong nonlinear interactions. Optically-pumped polariton lasers and polariton Bose-Einstein condensates in vertical microcavities have been demonstrated in a broad panel of semiconducting materials, from GaAs \cite{bajoni_polariton_2008, Butte_Transition_2002} and CdTe \cite{dang_stimulation_1998,kasprzak_bose-einstein_2006} for operation at cryogenic temperatures, to ZnO \cite{li_excitonic_2013}, GaN \cite{christmann_room_2008}, perovskites \cite{su_room-temperature_2017} and organic molecules \cite{kena-cohen_room-temperature_2010} for room temperature operation. The precise control over such polariton condensates has led to the observation of superfluidity \cite{amo2009superfluidity}, bright/dark spatial \cite{Walker2015, Sich2012, Walker2017} and temporal solitons \cite{Walker2015} and first signs of the polariton blockade regime  \cite{Munoz_Matutano_2019,Delteil_2019}; all these phenomena are based on strong polariton-polariton interactions at the heart of the so-called ``quantum fluids of light'' concept \cite{carusotto2013quantum}. Beyond single polariton condensates, the generation of multiple condensates with controlled interactions allows realizing advanced simulation schemes, such as polariton simulators~\cite{berloff2017realizing} and neuromorphic computing~\cite{ballarini2020polaritonic,kavokin2022polariton}. Patterned cavities have also allowed the first demonstration of topological lasing~\cite{st2017lasing}.

Waveguide polaritons have recently emerged as an alternative geometry to vertical microcavities thanks to the inherently long lifetime ~\cite{mechin2025time} and relatively simple fabrication. 
Recently, optically-pumped waveguide polariton lasers \cite{jamadi_edge-emitting_2018,Suarez-Forero_Electrically_2020,Souissi_Ridge_2022,Delphan_Polariton_2023,Oliveira_Whispering_2024} compatible with standard photonic integrated circuits have been demonstrated. Furthermore, the external control of nonlinearities therein \cite{Liran2025}, and the combination of multiple optical signals within a polariton node \cite{Han_2024}, now picture a complete platform for integrated polariton-based devices. Thus, overall the only missing element enabling a full-polaritonic integrated photonics platform is an electrically-injected polaritonic source compatible with such geometry. Early demonstrations of optically-pumped waveguide polaritons were based on GaAs, which only operate at temperatures below 50K \cite{Suarez-Forero_Electrically_2020}, and on ZnO \cite{jamadi_edge-emitting_2018}, which is not suitable for electrical injection. Perovskites and GaN-based devices presently appear as the most promising platforms capable of sustaining polaritons up to room temperature and displaying the necessary technology for electrical injection.  In particular, the fabrication of GaN polariton waveguides can take advantage of the 30 years long development of nitride-based optoelectronics, including commercial blue to ultraviolet light emitting diodes and laser diodes, as well as nitride-based integrated photonic devices \cite{jung_green_2014,trivino_continuous_2015,Liu_Ultra_2018,Siddharth_ultraviolet_2022,Wunderer_Single_2023}. Compared to commercial GaN visible lasers based on InGaN/GaN quantum wells, GaN polariton lasers need to be specifically designed to preserve excitons strongly coupled to guided photons under operation conditions, which is eased in bulk GaN material. Indeed, optically-pumped GaN waveguide polariton lasers have been demonstrated, the operation regime, continuous-wave \cite{Souissi_Ridge_2022,Delphan_Polariton_2023} or pulsed mode-locked \cite{Souissi_Mode_2024}, being dependent on the polaritonic gain bandwidth and relaxation efficiency. 

Shifting from optical pumping to electrical injection is a key towards the development of scalable polaritonic devices. The first electricallly-injected semiconductor polariton laser was demonstrated in a GaAs-based microcavity geometry at liquid-helium temperature under a strong magnetic field~\cite{schneider_electrically_2013,Gagel_2024}. Perovskite-based devices presently require a genuine cooling and specific injection protocol in order to obtain polariton lasing under electrical injection at $T=8$~K \cite{Pushkarev_2025} or additional optical pumping~\cite{elkhouly2024electrically}.

In this work, we rigoroulsy demonstrate the operation of a GaN polariton laser at room temperature under electrical injection, relying on a scalable micro-nanofabrication process. Its ridge geometry, very similar to commercial GaN-based lasers and in which the GaN waveguide core is inserted into an epitaxial p-i-n junction, is adapted to integrated photonic platforms. The strong coupling between excitons and photons under electrical injection of carriers is probed by the cavity free spectral range (FSR), which provides the dispersion of the polariton mode. The specificities of such a polariton laser are demonstrated: compared to lasers relying on electron-hole gain, the absence of reciprocity between absorption and stimulated emission here allows for a short injection length covering just a fraction of the total laser length, which would prevent lasing in standard lasers. This breakdown of reciprocity enables too the observation of Fabry-Perot modes far below threshold and, thus, the continuous monitoring of the polariton dispersion across the polariton lasing threshold; the large polaritonic gain allows cavity lengths, e.g. (60~µm), much shorter than in conventional lasers, enabling to reduce the lasers footprint. Finally, polariton laser devices from the same wafer but without the electrical contacts were operated under optical pumping, achieving polariton lasing and reproducing previously reported spectral features, confirming that in our design the exciton-photon coupling is not affected by the doped semiconductor layers.

\section*{Results}\label{sec2}

\begin{figure}[tbhp]
\centering{\includegraphics[width=15cm]{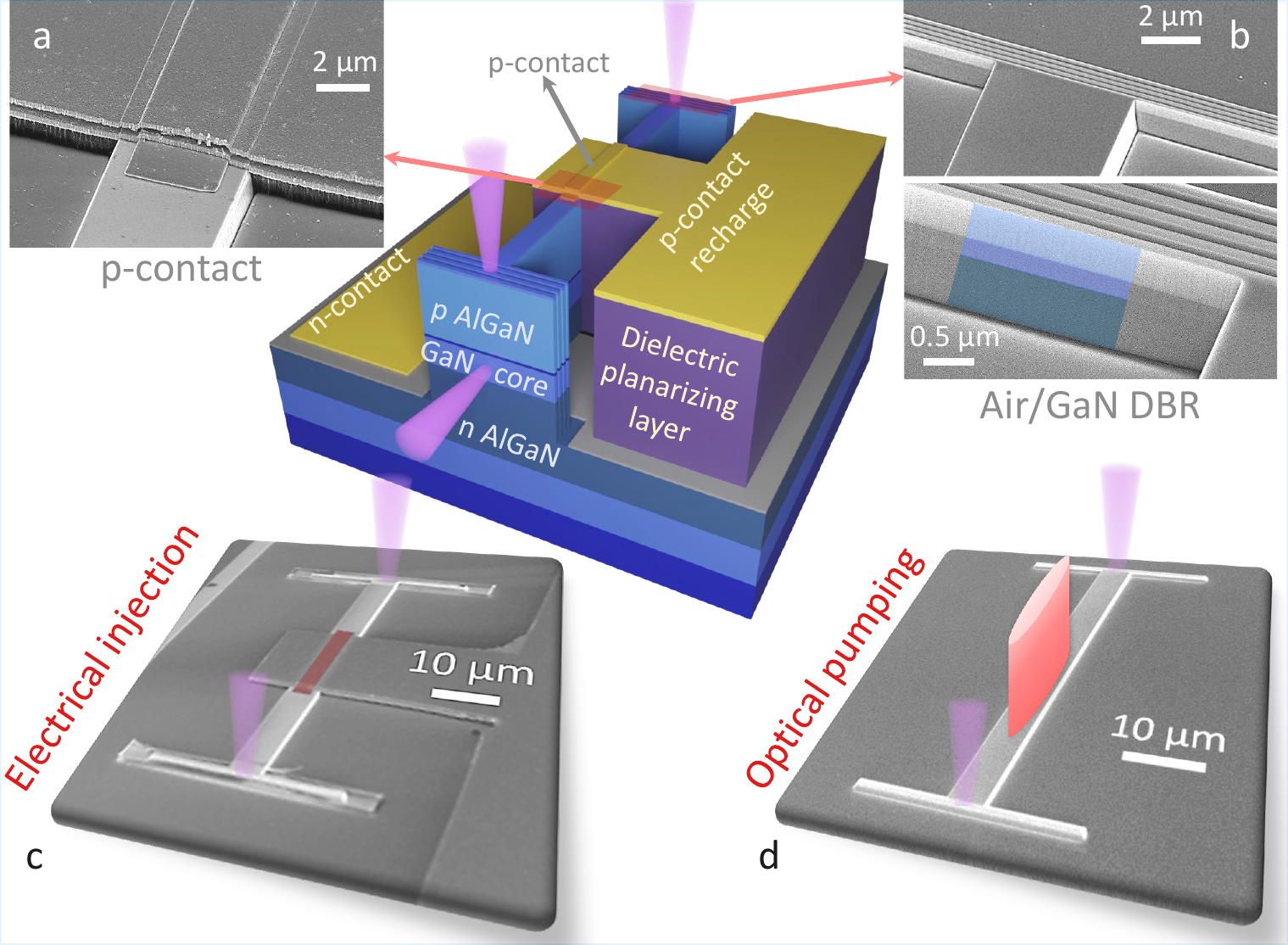}} 
\caption{GaN-based polariton ridge waveguide device. The schematic illustrates the layer stack, including the ridge, the air/GaN distributed Bragg reflector, and a planar p-contact. Scanning electron microscopy (SEM) images highlight (a) the p-contact covered with the p-recharge pad , (b) the DBR , (c) the fabricated 60~µm-long ridge waveguide with a 20~µm-long p-contact area in its center and terminated by the vertical DBRs designed for electrical injection, and (d) the fabricated 60~µm-long cavity designed for optical pumping.}
  \label{fig:sample}
\end{figure}

A schematic of the GaN-based polariton laser diode (LD) is shown in Fig.~\ref{fig:sample}. The planar waveguide is grown on a free-standing GaN substrate, with the GaN core  hosting the exciton-polaritons sandwiched between p- and n-doped AlGaN top and bottom claddings, respectively, forming a double heterostructure p-i-n junction. High-reflectivity distributed Bragg reflectors (DBRs), which act as reflectors but also as light outcouplers, are integrated at both ends of the ridge waveguide and define the laser Fabry-Perot cavity (Fig.~\ref{fig:sample}b). Two sets of identical cavities, with and without metal contacts, are processed on the same wafer for comparing quantitatively electrical injection and optical pumping, respectively (Fig.~\ref{fig:sample}c,d). The details on the epitaxial design and fabrication process are presented in the Appendix~\ref{app:Epitaxial design and fabrication details}. 

\begin{figure}[tbhp]
\centering{\includegraphics[width=15cm]{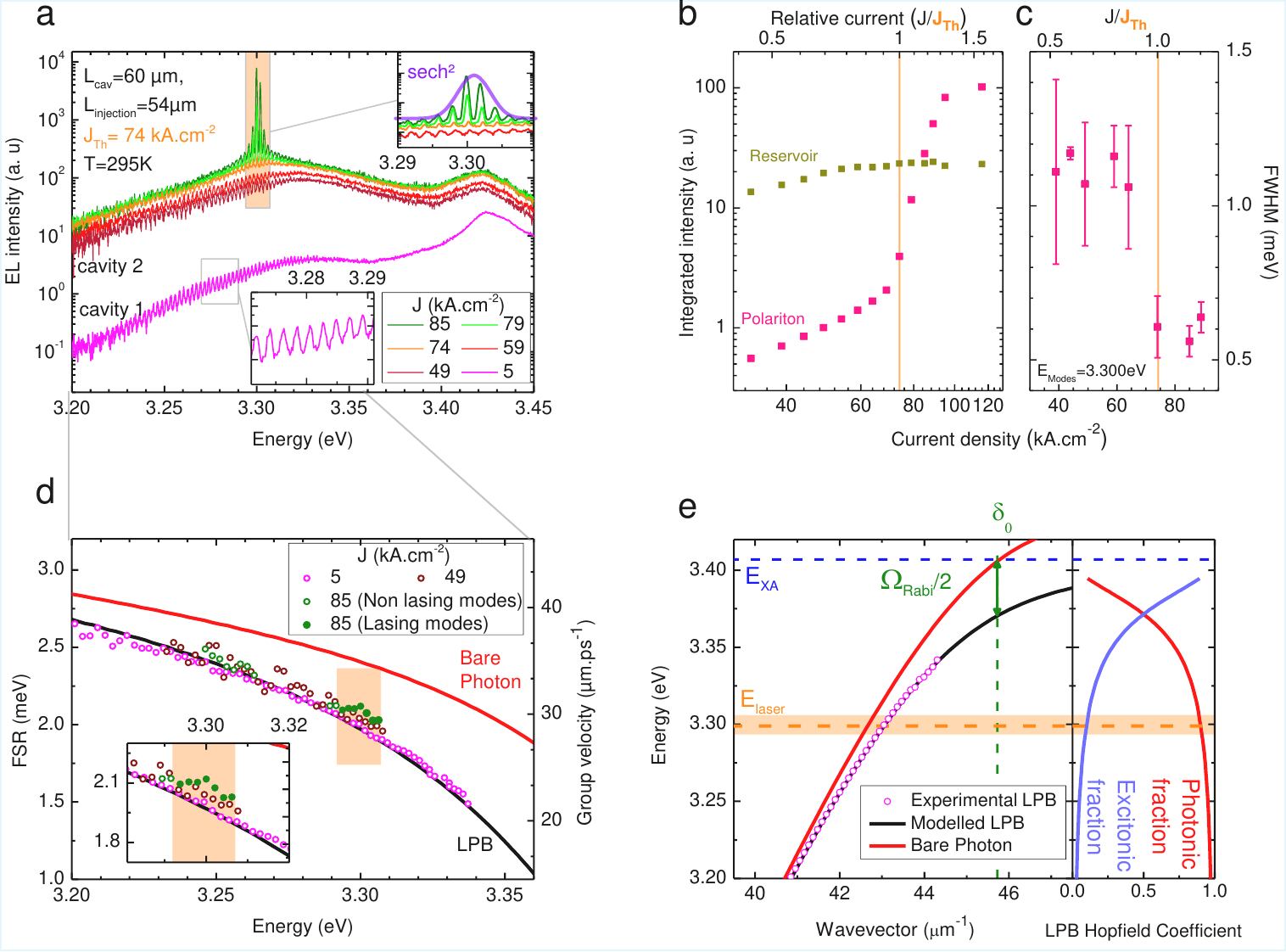}} 
  \caption{(a) Emission of 60~µm-long electrically injected cavities; inset: associated $\sech^{2}$ fit (pink line) of the lasing envelope, (b) Integrated intensity of the excitonic reservoir measured near the contact between 3.36 and 3.45~eV, and of the lasing modes measured at the DBR between 3.294 and 3.306~eV indicated by the orange rectangle; (c) Spectral narrowing of the central lasing mode at 3.300~eV (Lorentzian fit), (d) Experimental FSR (open dots: non lasing modes, full dots: lasing modes) and calculated FSR from the lower-polariton-branch (LPB) (black line) and bare photon dispersion (TE0 mode, red line) without exciton, (e) Simulated waveguide dispersions (Eigenmode expansion method, plain lines) and experimental dispersion  at room temperature [right], Hopfield coefficient Excitonic (light blue line) and photonic (red line) of the LPB [left]}
  \label{fig:Emission spectrum}
\end{figure}

In a conventional semiconductor laser diode the injected carriers fill the conduction and valence bands so that, by fulfilling the condition of population inversion \cite{Bernard_Laser_1961}, optical gain can be promoted. In the present device injected carriers bind together and form excitons, which can be observed under the thin contact in reflectivity at the energy of the free exciton $E_{XA}$ (Appendix~\ref{app:Exciton energy in the GaN core layer}, Fig.~\ref{fig:Reflectivity}). These excitons form a dense reservoir rather than an electron-hole plasma, strongly couple to the waveguide photonics modes to form polaritons, relax in energy along the lower polariton branch and populate the quantized Fabry-Perot polariton modes. These polariton modes, which are extracted and measured at the location of the DBRs, are clearly visible not only at high current densities but also at current densities much lower than the polariton lasing threshold current (Fig.~\ref{fig:Emission spectrum}a). As the current is increased, the population of one of the Fabry-Perot polariton modes exceeds unity, triggering the stimulated scattering of polaritons and enabling an efficient polariton relaxation from the exciton reservoir to the lasing polariton mode at 3.29~eV.

The electroluminescence (EL) spectrum of a 60 µm-long cavity evidencing such a polariton laser effect is presented in Fig.~\ref{fig:Emission spectrum}a
, with a threshold current density J$_\text{th}=$~74~kA.cm$^{-2}$. As we will discuss later the lowest threshold current density we could achieve, on a longer cavity, was 18~kA.cm$^{-2}$.
The EL-current characteristic of the polariton laser is shown on Fig.~\ref{fig:Emission spectrum}b. The intensities of the exciton reservoir and of the polariton emission at the lasing energy evidence two important features: first, the reservoir intensity increases with the current below threshold and then saturates, evidencing the clamping of the reservoir density; second, the emission intensity of the polariton laser displays a strong non-linear increase of almost 2~orders of magnitude at threshold, with a corresponding spontaneous emission factor $\beta$ around $0.02$. Simultaneously, Fig.~\ref{fig:Emission spectrum}c, the polariton lasing mode (at 3.3~eV) linewidth narrows down from 1.1 to 0.6~meV, manifesting the onset of temporal coherence.

To  demonstrate unambiguously that the current laser operates in the strong-coupling regime, we need to measure the polariton energy dispersion under electrical injection below and above the lasing threshold and evidence the anti-crossing between the guided photon mode and the exciton, which is the smoking gun of strong-coupling. To do so we measure the energy dependence of the cavity free spectral range (FSR), which is proportional to the first derivative of the polariton dispersion ($FSR=(L_{cav}/\pi)(\partial E_{LPB}/\partial \beta)$). At the lowest current density (5 kA.cm$^{-2}$, cavity~1), the Fabry-Perot modes are visible over a broad spectral range as shown in the inset of the Fig.~\ref{fig:Emission spectrum}a. The experimental FSR and the corresponding dispersion (Fig.~\ref{fig:Emission spectrum}d,e, open dots) are reproduced by the simulated polariton dispersion with a Rabi splitting $\Omega_{Rabi}$=71 $\pm$ 10 meV, much larger than $k_B T$, evidencing the formation of robust polaritons within the structure. The modeling of the polariton dispersion also provides the Hopfield coefficients of the lasing modes, whose exciton fraction is estimated to 11$\%$ (Fig.~\ref{fig:Emission spectrum}e). 

It is interesting to note that the polaritonic dispersion remains almost unchanged across the lasing threshold, both for polaritons away from the lasing mode (modes at 3.25~eV) as well as for polaritons close to the lasing modes (around 3.29~eV). Most importantly, and consistent with the polaritonic nature of the lasing modes, the blueshift of each Fabry-Perot mode for current densities above threshold is in the meV range. This is much smaller than the 15-20 meV energy difference between the polariton and the bare photon dispersions at the lasing energy, as shown in Fig.~\ref{fig:Emission spectrum}e, confirming that polaritons survive across the lasing threshold. This blueshift is induced by the interactions between polaritons and/or with the excitonic reservoir. Under the current operation conditions, it leads to a flat energy dependence of the FSR at threshold for the modes involved in lasing, i.e. to a linearized polariton dispersion (Fig.~\ref{fig:Emission spectrum}d). Such renormalized dispersion, which allows the onset of resonant parametric oscillation processes, and the $\sech^{2}$ spectral profile of the laser emission shown on the inset of Fig.~\ref{fig:Emission spectrum}a indicate that the our GaN polariton lasers operate in the mode-locked regime, as discussed in detail in \cite{Souissi_Mode_2024} for a polariton laser under optical injection. 

\begin{figure}[tbhp]
\centering{\includegraphics[width=7cm]{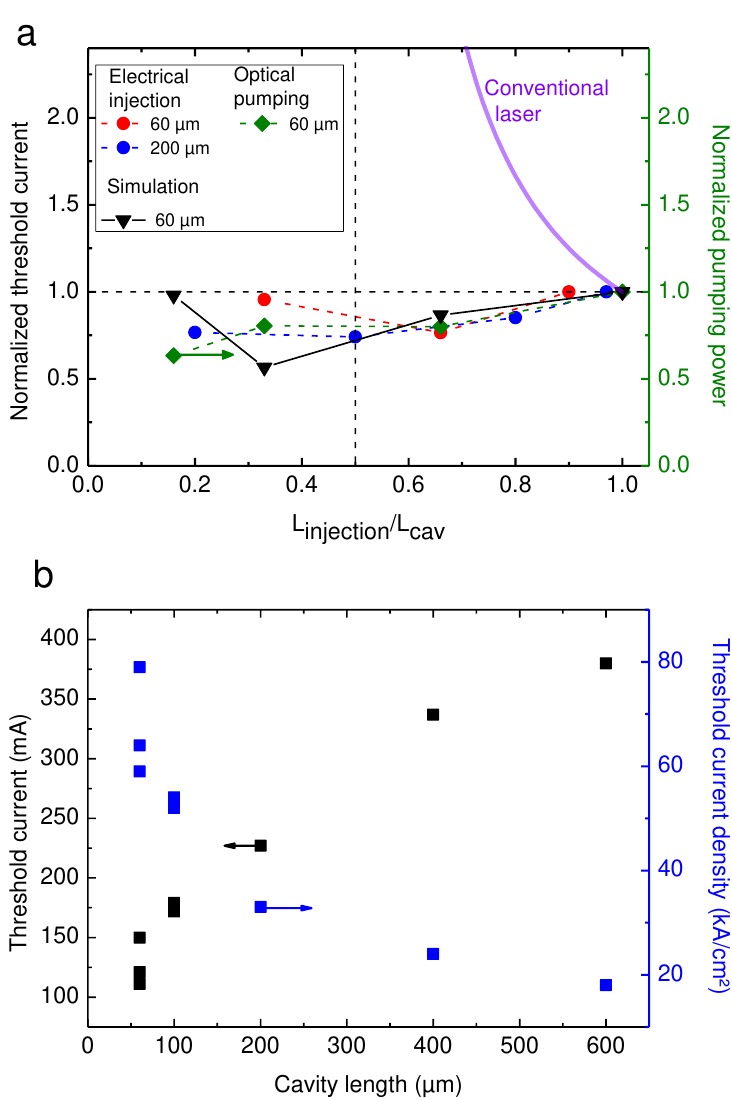}}
  \caption{(a) Normalized threshold current (respectively normalized pumping power) at laser threshold as a function of the injection-to-cavity ratio (L$_\text{injection}/$L$_\text{cav}$), normalized to its value for L$_\text{injection} \simeq $~L$_\text{cav}$, for electrically injected LDs (60~µm and 200~µm cavities) and optically pumped 60~µm cavity. (b) Threshold current (black squares, left axis) and threshold current density (blue squares, right axis) as a function of cavity length. }
  \label{fig:Laser threshold}
\end{figure}

A key feature of waveguide polariton lasers, which differentiates them from standard electron-hole plasma edge-emitting lasers, lies in the singular dependence of their lasing threshold as a function of the injection-to-cavity lengths ratio. In conventional lasers, the threshold increases drastically as soon as a small section of the cavity is not pumped. More precisely, the reciprocity between absorption and gain implies that conventional lasing can not be achieved when less than 50$\%$ of the cavity is electrically (or optically) pumped, given that in such situation the absorption of the unpumped region would necessarily override the gain provided by the pumped region.
In polariton lasers, where this reciprocity condition is absent, the behavior is radically different, as absorption at the polariton lasing energy is only residual and could be, in principle, brought to zero in an ideal media at 0K. As a result, even in real  polariton lasers (i.e. displaying some residual absorption) one can expect to attain the polariton lasing threshold even when pumping a small fraction of the total cavity length, as demonstrated recently in optically-pumped waveguide polariton lasers~\cite{Souissi_Ridge_2022} where only 15$\%$ of the cavity length was excited. In the current work, electrically-injected cavities with different total lengths of 60 and 200~µm were fabricated, and contacts of varying length covering from 20$\%$ to 97$\%$ of the cavity length were deposited. Electrically-injected polariton lasing at room-temperature was achieved in all cavities and, in particular, in the cavities for which the pumped region is smaller than half of the cavity length, as displayed in Fig.~\ref{fig:Laser threshold}a. Noteworthly, when the lasing threshold, measured for a fixed cavity length but different injection-to-cavity lengths ratio under electrical injection (or optical pumping), is normalized by the lasing threshold measured for a pumped area covering the whole cavity length (L$_\text{inj}=$ L$_\text{cav}$), it is observed that the set of data obtained for different cavity lenghts (e.g. L$_\text{cav}=$~200~µm, 60~µm) follow the same trend and reasonably overlap. In contrast, the expected trend for a conventional laser diode of comparable geometry would be very different as highlighted in Fig.~\ref{fig:Laser threshold}a, and further detailed below.


The overall process of electrically-injected polariton lasing involves several distinct physical process that occur sequentially: carrier injection from spatially distant electrical contacts, formation of excitons and the associated exciton reservoir, polariton relaxation along the polariton dispersion including stimulated scattering and, finally, polariton lasing. To get deeper insight into this sequence of events we performed numerical simulations of electrical injection, exciton formation and polariton relaxation using a combination of NextNano \cite{Trellakis_2006} and semi-classical Boltzmann equations, as described in detail in~\cite{malpuech2002room,solnyshkov2009theory} and employing physical parameters of GaN polaritonic waveguides taken from the most recent measurements~\cite{mechin2025time} (see Appendix~\ref{app:Theory} for more details). Our simulations show that the carrier density in the waveguide at the threshold currents is of the order of $5\times 10^{17} \text{cm}^{-3}$, compatible with the strong coupling regime. Furthermore, the threshold current densities we deduce from the simulations are comparable with experiments. The numerical results for the normalized threshold current for different injection-to-cavity lengths ratios are shown in Fig.~\ref{fig:Laser threshold}a. Additional simulations for larger cavities have shown a threshold current density reduction by a factor $1.9$ between L$_\text{cav}=$~60~µm and L$_\text{cav}=$~200~µm, also compatible with the experimental results (Fig.~\ref{fig:Laser threshold}b).


\begin{figure}[tbhp]
\centering{\includegraphics[width=7cm]{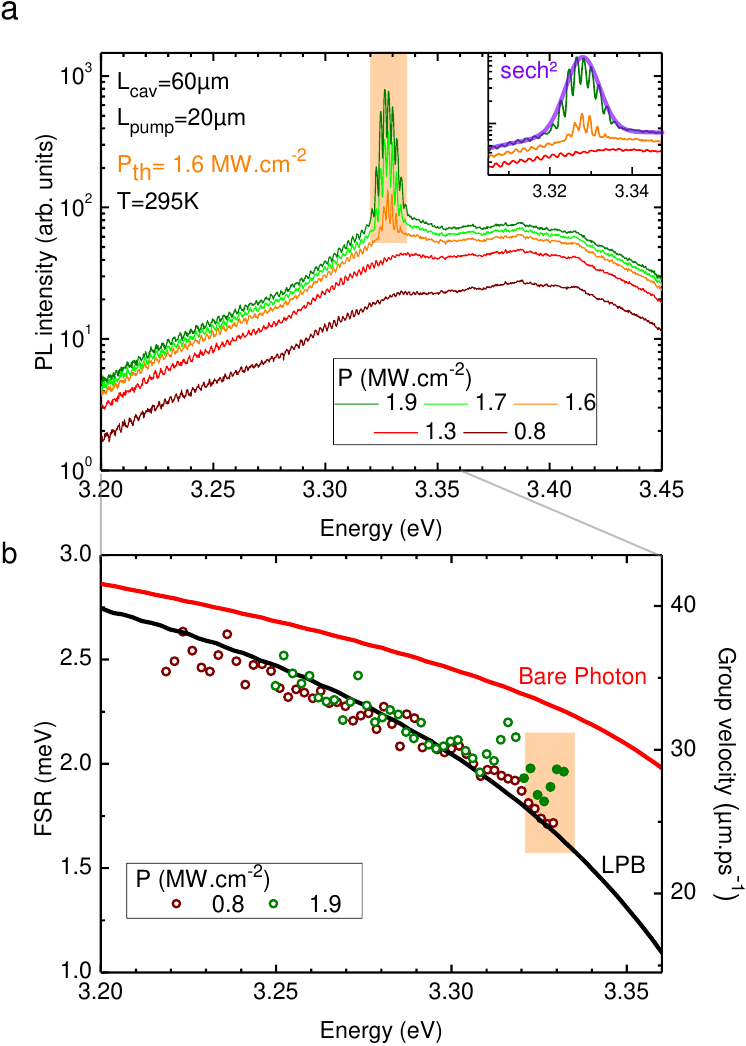}} 
  \caption{(a) Power dependant emission of a 60µm-long under optical pumping and the associated $\sech^{2}$ fit (pink line) of the lasing envelope [inset], (b) Experimental FSR (dots) and calculated FSR from the LPB (black line) and bare photon mode (TE0 mode, red line)} 
  \label{fig:Optical_pumping}
\end{figure}

Interestingly, the electrical and optical injection schemes can be directly benchmarked in the present polariton waveguide design thanks to the easy optical (and electrical) access to the active region (see Appendix~\ref{app:Epitaxial design and fabrication details}), which is less straightforward in vertical microcavities due to the presence of the top DBR.
The emission, collected at one of the DBRs position, of an optically-pumped 60~µm cavity as a function of pumping irradiance is shown in Fig.~\ref{fig:Optical_pumping}a. The spectral features are very similar to the ones measured under electrical injection: Fabry-Perot modes can be clearly resolved below threshold and the polariton dispersion reproduces the one measured under electrical injection, with a similar excitonic fraction of 16$\%$ for the lasing modes. At threshold, the same nonlinear intensity increase of a few modes following a $\sech^{2}$ envelope is observed. Moreover, the normalized pumping power at threshold P$_\text{th}$ follows a similar dependence vs the injection-to-cavity lengths ratio as the threshold current density for the electrically-injected devices (Fig.~\ref{fig:Laser threshold}a).  The pump power at threshold is similar to previous devices without doped layers \cite{Souissi_Mode_2024}, showing that the losses related to the doped claddings are small compared to other losses in the cavity. The laser energy of electrical-injected is 30~meV smaller, consistent with a smaller average exciton thermal energy in the reservoir under electrical injection \cite{solnyshkov2009theory}.

Let us finally compare the coupling regime and the threshold current density of the current polariton laser devices with conventional diode lasers based on nitride semiconductors. In nitride-based microcavities, the strong coupling regime and the polariton laser were first demonstrated with bulk GaN \cite{christopoulos_room-temperature_2007}, as in the present work, and soon after with GaN/AlGaN quantum wells \cite{christmann_room_2008,levrat_condensation_2010}. In contrast, InGaN/GaN QWs used in conventional visible laser diodes are highly sensitive to alloy fluctuations of the active layers, which induce a significant inhomogeneous broadening, hindering the formation of polaritons within structures completely equivalent to those used in commercial InGaN/GaN laser diodes \cite{Glauser2014}. Regarding the threshold current, conventional (commercial) ridge lasers usually embed much longer cavities than the present devices, and typically do not operate for cavities shorter than 200~µm. Fig.~\ref{fig:Laser threshold}b displaying results achieved from cavity length ranging from 60~µm to 600~µm, can be used to benchmark the threshold current polariton lasers with respect to conventional InGaN/GaN laser diodes. The threshold current density J$_\text{th}$ is reduced by a factor 4, from 79 to 18~kA.cm$^{-2}$, as the cavity length L$_\text{cav}$ increases from 60~µm to 600~µm. State-of-the-art threshold current densities in  InGaN QWs lasers reach values of few kA.cm$^{-2}$ \cite{Farrell2010, Zhang2023,Kawaguchi2016}, up to 10~ kA.cm$^{-2}$ for a 45~µm-long cavity \cite{Zhang2019}. This is less than one order of magnitude lower than the current densities of our electrically-injected polariton lasers, which are not yet optimized and which are based on a bulk GaN active region. Future work, based on cavities employing more similar active regions, will enable an even more direct comparison of the thresholds for both lasing mechanisms.

\section*{Conclusion}

To summarize, we have presented an electrically injected GaN polariton laser operating at room-temperature with a threshold current density J$_\text{th}$=74~kA.cm$^{-2}$ (resp 18~kA.cm$^{-2}$) for a 60~µm-long (resp 600~µm) laser device.
The specificities of waveguide polariton lasers are evidenced, with a polariton dispersion displaying the anticrossing associated to the strong coupling between excitons and photons, a large available gain, and the possibility to achieve lasing with short injection sections without being limited by the transparency condition associated to population inversion. Electrically-injected devices are directly compared with optically-pumped ones operating in strong-coupling regime, showing important similarities. The room temperature operation, the similarity with the design of conventional nitride lasers, and the use of a saclable technology, represent significant advantages for the integration of polariton lasers in photonic integrated circuits. Indeed, first reports of electrically-injected polariton lasers required the use of cryogenic temperatures, because of the small exciton binding energy in GaAs-based heterostructures, and a strong magnetic field \cite{schneider_electrically_2013, Gagel_2024}, or could just occur at uncontrolled polariton traps, as recently reported in a perovskite microcavity \cite{Pushkarev_2025}.
Additionally, the combination of the polaritonic non-linearity and the exciton dispersive effects results in the mode-locked regime of the laser. These structures highlight the potential of such devices for future integration of polaritonic devices, with strong optical nonlinearities and a small footprint.

\backmatter

\bmhead{Funding}

This work was supported by European Union's Horizon 2020 program, through a FET Open research and innovation action under the grant agreement No. 964770 (TopoLight), and by the European Union EIC Pathfinder Open project “Polariton Neuromorphic Accelerator” (PolArt, Id: 101130304). Additional support was provided by the ANR Labex GaNext (ANR-11-LABX-0014), the ANR program "Investissements d'Avenir" through the IDEX-ISITE initiative 16-IDEX-0001 (CAP 20-25), the ANR project MoirePlusPlus, the ANR project "NEWAVE" (ANR-21-CE24-0019), and the R\'egion Occitanie. C2N is a member of RENATECH-CNRS, the French national network of large micro-nanofacilities.

\bmhead{Conflict of interest/Competing interests}
The authors declare no competing interests.

\bmhead{Authors' contributions}

H.S. and V.D. have contributed equally to the present work.
\\Device processing: H.S., M.G., E.C., S.C., M. A., M.Z., J.Z. and S.B. Device characterization: H.S., V.D., C.B., L.D., S.C., M.A., S.B. and T.G.. Design and growth : B.A., E.F. and J. Z.. Simulation and Theory: J.Y.D., G.M. and D.D.S.. Writing-original draft: H.S., V.D. and T.G.. Writing-review and editing: all the authors. 

\newpage

\section*{Appendix}

\appendix

\section{Epitaxial design and fabrication details}\label{app:Epitaxial design and fabrication details}
\begin{figure}[h]
\centering{\includegraphics[width=14cm]{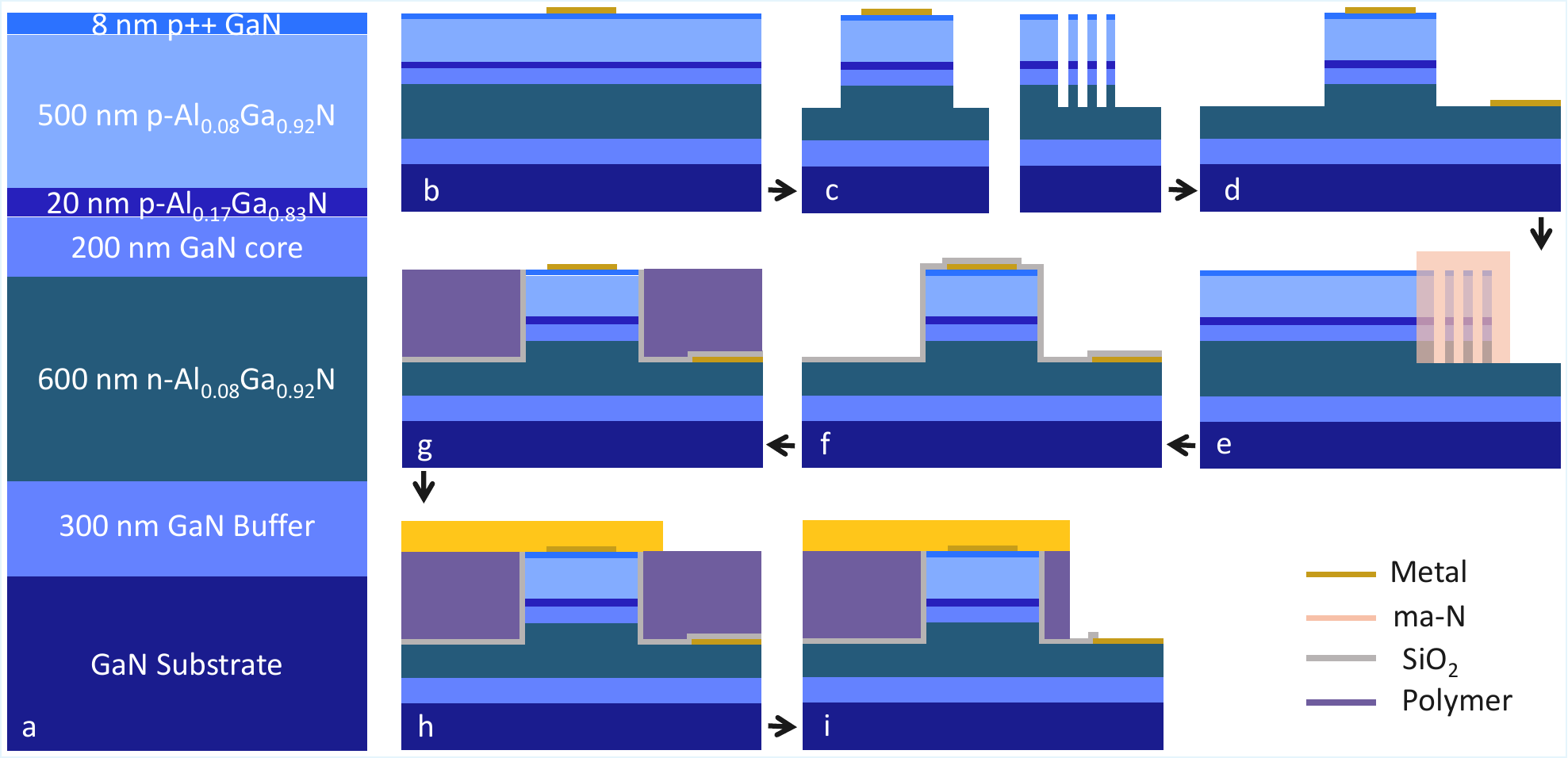}} 
  \caption{(a) Schematic cross-sectional view of the as-grown LD structure. (b) Marks and p-contact deposition. (c) Ridge and DBR deep etching down to the n-type AlGaN. (d) n-contact pad deposition. (e) ma-N resist on DBR for protection. (f) $\text{SiO}_{2}$ insulation layer deposition. (g) Polymer planarization and opening. (h) p-contact pad deposition. (i) Polymer opening on n-contact pad.}
  \label{fig:Fabrication process}
\end{figure}

The sample was grown by metal-organic chemical vapor deposition (MOCVD) on a 2-inch commercial free-standing GaN substrate. The epitaxial heterostructure is schematically shown in Fig.~\ref{fig:Fabrication process}a. After the deposition of a 300 nm-thick GaN buffer layer, a 600 nm-thick n-type $\text{Al}_{0.08}\text{Ga}_{0.92}\text{N}$ cladding layer (Si-doped, concentration 5.10$^{18}$ $cm^{-3}$) was grown. The active region consists of a 200 nm-thick GaN waveguide layer. It is followed by a 20 nm-thick p-type AlGaN layer (Mg-doped, concentration 10$^{19}$ $cm^{-3}$ ) containing 17$\%$ of Al , acting as an electron-blocking layer (EBL). A 500 nm-thick p-type $\text{Al}_{0.08}\text{Ga}_{0.92}\text{N}$ layer (Mg-doped) is then grown to serve as the upper p-cladding. Finally, the structure is capped with an 8 nm-thick highly Mg-doped GaN layer to facilitate ohmic contact formation. 

Prior to laser diode (LD) fabrication, the sample was cleaned using sequential rinses in acetone and isopropanol (IPA) followed by a 5 min immersion in a piranha-like solution  ($\text{H}_{2}\text{SO}_{4}$ – 20 vol. / $\text{H}_{2}\text{O}_{2}$ – 6 vol. / $\text{H}_{2}\text{O}$ - 25 vol.) to remove organic residues. After rinsing in deionized water (DIW), the sample was treated for 3 min in an HCl-based solution (2 vol. of DIW / 1 vol. of 37$\%$ HCl) and rinsed again to remove native oxides before proceeding to the first lithography step.

The LD fabrication process involves six electron-beam lithography steps. In the first step, the p-type contact consisting of a  Ni(20~nm)/Au(200~nm) metal stack is defined with various lengths by electron-beam-vacuum evaporation and metal lift-off (Fig.~\ref{fig:Fabrication process}b). Right prior loading the sample into the evaporation chamber, it is immersed for 3~min in the HCl-based solution, followed by a short rinse in DIW to remove interfacial native oxides. The in-plane laser cavities consisting of a 5-µm wide waveguide (with the p-contact on top) and terminated by 4-pair ($\lambda$/4 in air - 3$\lambda$/4 in semiconductor) vertical DBRs are subsequently defined by chlorine-based inductively coupled plasma reactive ion etching (ICP-RIE) (Fig.~\ref{fig:Fabrication process}c). After this etch down to 1.1~µm to access the n-type cladding layer , the n-type contact pad is defined by deposition of a Ni(20~nm)/Au(200~nm) metal stack (Fig.~\ref{fig:Fabrication process}d). Rapid thermal annealing (RTA) of the contacts is then performed at 450°C for 4 min under a pure oxygen ambient, enabling a reduction of the Schottky barrier height and improved ohmic behaviour of the contacts~\cite{Duraz_2026}. The DBRs, are then temporarily covered with ma-N resist using a 4$^\text{th}$ e-beam lithography step for their protection during subsequent processing which is followed by the sputtering of a 50 nm-thick $\text{SiO}_{2}$ layer for electrical insulation (Fig.~\ref{fig:Fabrication process}e, \ref{fig:Fabrication process}f). Ridge planarization is achieved by spin-coating then cross-linking an organic dielectric layer (SU-8 negative-tone epoxy type resist) and controlled etching of the polymer and $\text{SiO}_{2}$ layers by capacitively coupled plasma reactive ion etching (CCP-RIE) to expose the surface of the p-contact on top of the ridge mesa (Fig.~\ref{fig:Fabrication process}g). A p-contact recharge pad is then defined by deposition of a Ni(20~nm)/Au(200~nm) stack, subsequently thickened using a Cu(300~nm)/Ti(20~nm)/Au(200~nm) metal stack (Fig.~\ref{fig:Fabrication process}h). This allows the formation of a thick and flat p-contact pad (see Fig.~\ref{fig:sample}c, \ref{fig:sample}d) so as to reduce localized thermal and electrical stress during current injection. The final lithography step consists in etching the remaining polymer and $\text{SiO}_{2}$ layers by CCP-RIE to open access to the n-contact pad (Fig.~\ref{fig:Fabrication process}i). The sample is then immersed in a NMP (N-Methyl-2-Pyrrolidone)-based stripper (Remover PG) at 70°C, followed by IPA rinses to remove the ma-N resist from the DBRs. Finally, ridge-waveguide cavities were also processed from the same sample, without any p-contact evaporated on the top of the ridge neither p-contact recharge, allowing for tests under optical pumping and direct comparison between the electrical injection and optical pumping schemes.

\section{Spectroscopy under electrical injection and optical pumping}\label{app:Spectroscopy under electrical injection and optical pumping}
LD devices were tested at room temperature. Under continuous-wave (CW) operation, current–voltage (I–V) characteristics were recorded over a voltage range of -2~V to 4~V to evaluate possible current leakage. Micro-Electroluminescence (µEL) imaging is performed under quasi-CW electrical injection (5~kHz repetition rate, 0.05$\%$ duty cycle, nominal pulse width 100~ns). The oscilloscope time-domain measurements indicate that the injected current amplitude is slightly lower than the nominal current value. The current in the presented datas correspond to the actual peak value of the current pulse, while the current densities correspond to the current peak value divided by the length of the p-contact pad (on top of the ridge). The emission is collected using a 10x microscope objective and analyzed with a spectrometer (Horiba iHR550) equipped with a charge-coupled device (CCD) camera for real-space imaging. This configuration enables clear discrimination between emission from the excitonic reservoir near the p-contact, and guided polariton emission outcoupled at the DBRs (Fig.~\ref{fig:Experimental setup}).

Under optical pumping, the cavity is excited at its centre using a pulsed laser source (355~nm wavelength, 7~kHz repetition rate, 4~ns pulse width) resonantly exciting the excitonic reservoir. The beam is shaped into a line of adjustable length $L_\text{injection}$ with a cylindrical lens, and focused at the center of the cavity. This allows partial or complete excitation of the cavity, similarly to the series of electrically injected devices, so to control the size of the excitonic reservoir. The emission is collected with the same 10x microscope objective.

\section{Theory}\label{app:Theory}
The electrical injection simulations were performed using the NextNano software package~\cite{birner2007nextnano}. The simulations provide the carrier density in the waveguide region as a fucntion of injected current. We used the following key parameters determining the carrier lifetimes in the NextNano package: $\tau_n=\tau_p=1$~ns for the Shockley–Read–Hall (SRH)  mechanism and $b~=~5e^{-9}$~cm$^{3}$.s$^{-1}$, leading to a radiative lifetime of 190~ps over the guide. With these parameters, and for the injection current densities used, we find that the radiative decay dominates over SRH. The effective carrier lifetime in the center of the waveguide is of the order of 150~ps, in agreement with our recent experimental observations~\cite{mechin2025time}. 

The densities of carriers of both types were found to be comparable and in the range of 7e$^{17}$~cm$^{-3}$ at threhold. The subsequent simulations of exciton formation and polariton relaxation were based on the semi-classical Boltzmann equations combining real and reciprocal space resolution, described in detail in the references~\cite{malpuech2002room,solnyshkov2009theory,ciers2017propagating}:
\begin{eqnarray}
\frac{\partial n_{k,r}}{\partial t}&=&P_{k,r}-\Gamma_k n_{k,r}-n_{k,r}\sum\limits_{k'}W_{k\to k',r}\left(n_{k',r}+1\right)\nonumber\\
&+&\left(n_{k,r}+1\right)\sum\limits_{k'}W_{k'\to k,r}n_{k',r}-v_g(k)\frac{\partial n_{k,r}}{\partial r}
\end{eqnarray}
where $n_{k,r}$ is the exciton-polariton distribution function within real and reciprocal spaces. The decay rate $\Gamma_k$ accounts for radiative and non-radiative contributions for both photons and excitons (we were using the decay rates measured recently in~\cite{mechin2025time}). The pumping $P_{k,r}$ is computed from the carrier densities generated by NextNano. The scattering rates $W_{k\to k',r}$ account for exciton-exciton and exciton-phonon scattering mechanisms. The group velocity $v_g$ taken from the experimental dispersion describes the polariton propagation between the adjacent spatial cells.

\section{Experimental setup}\label{app:Experimental setup}

\begin{figure}[tbhp]
\centering{\includegraphics[width=12cm]{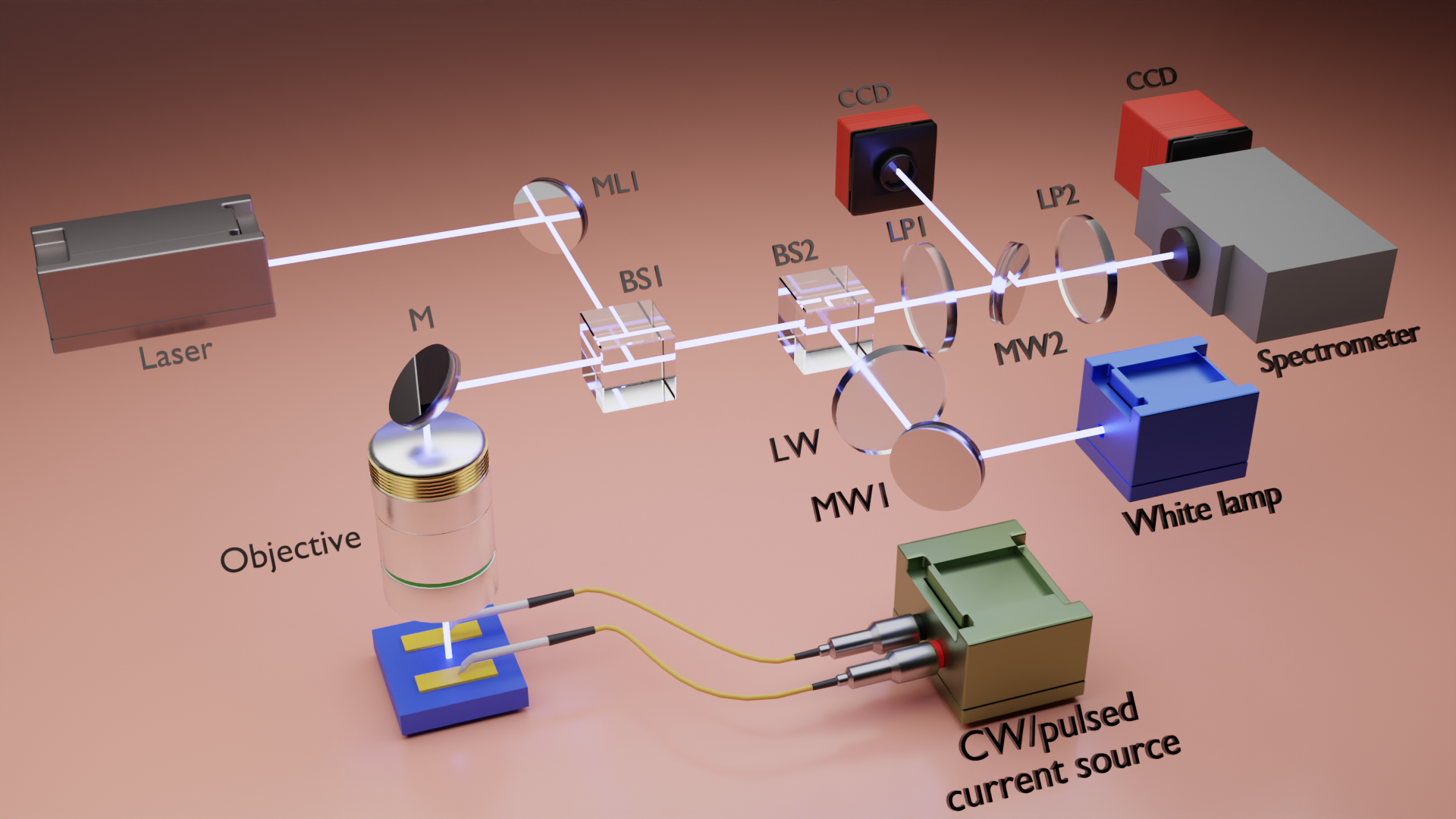}} 
  \caption{Experimental setup for photoluminescence and electroluminescence measurements. }
  \label{fig:Experimental setup}
\end{figure}

The experimental setup used for both µEL and µPL measurements is schematically depicted in Fig.~\ref{fig:Experimental setup}. Sample navigation is performed under white-light illumination provided by a white lamp. The light is focused onto the sample through a microscope objective, while the reflected signal is collected by the same objective. The collected light is relayed by lens LP1 and directed towards a CCD camera by inserting the removable mirror MW2, allowing real-time imaging of the sample surface.
\\ For µEL measurements, electrical excitation is supplied by a CW or pulsed current source connected to the device through electrical probes tips. The emitted light is collected by the same microscope objective and relayed by lens LP2 to image the cavity onto the spectrometer entrance slit. A charge-coupled device (CCD) mounted at the spectrometer output enables reconstruction of the real-space image.
\\ For µPL measurements, optical excitation is provided by a pulsed laser, while detection is performed using the same optical configuration as for µEL measurements.

\section{Imaging of the excitonic reservoir}\label{app:Imaging of the excitonic reservoir}

\begin{figure}[tbhp]
\centering{\includegraphics[width=14cm]{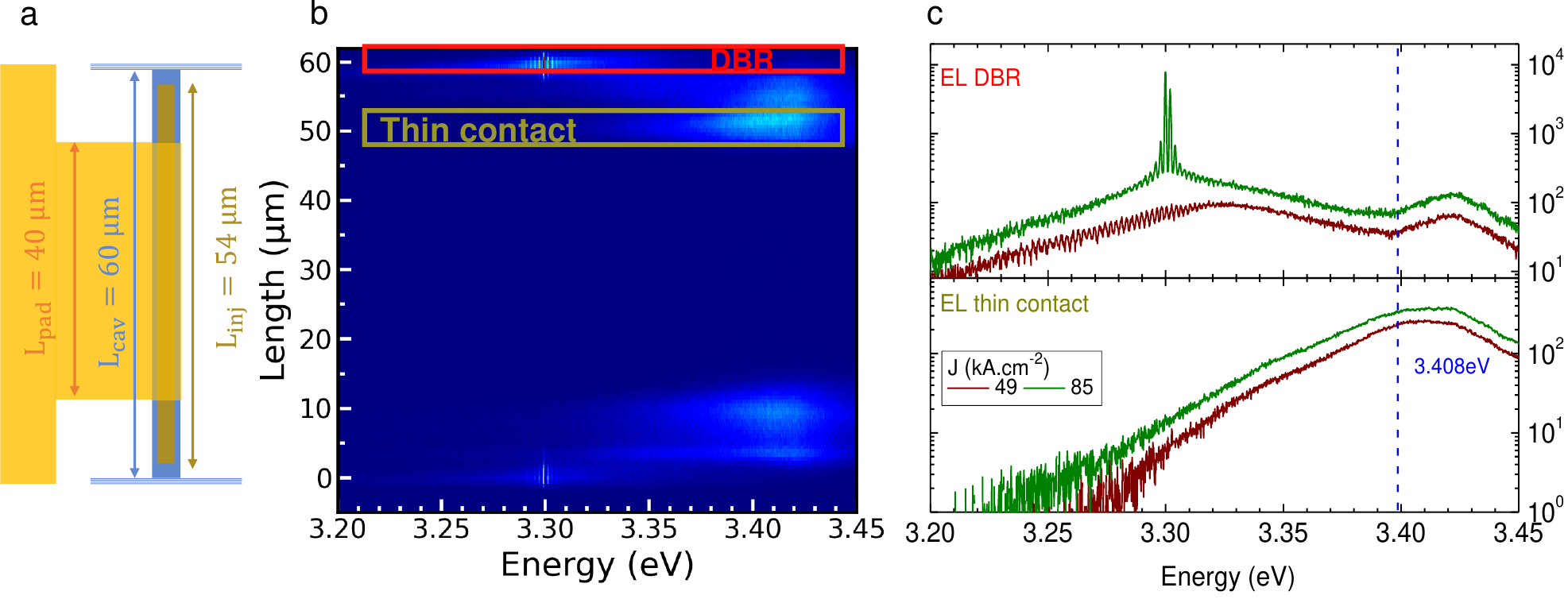}} 
  \caption{(a) Schematic of the cavity and the electrical contact (b) EL image of the cavity as a function of the energy and the position, (c) EL spectral profil at different position: at the DBR (top), under the thin contact (bottom)} 
  \label{fig:ResX}
\end{figure}

The excitonic reservoir is hosted in the GaN core layer of the laser device, and generated under the p-type Ni/Au contact. The Fig.~\ref{fig:ResX}a presents a scheme of the two-step contact deposition, illustrated in the SEM images (Fig.~\ref{fig:sample}a). The thick part of the pad is 40~µm-long and fully blocks the reservoir vertical emission, but the thinner 54~µm-long under-laying contact is only partially covering the waveguide. Therefore we can directly observe (i) the emission of the exciton reservoir in a short "thin-contact" section of the waveguide, and (ii) the polariton emission collected at the DBR position, as shown in the Fig.~\ref{fig:ResX}b. The corresponding spectra are presented in the Fig.~\ref{fig:ResX}c. Near the contact, the emission peaks between 3.408 and 3.42 eV and is attributed to the direct detection of the A and B excitons, whereas the two contributions from the reservoir and the LPB polaritons (below 3.35~eV) are observed. This allows the determination of the reservoir and polariton intensities shown in the Fig.~\ref{fig:Emission spectrum}b.

\section{Exciton energy in the GaN core layer}\label{app:Exciton energy in the GaN core layer}

\begin{figure}[tbhp]
\centering{\includegraphics[width=11cm]{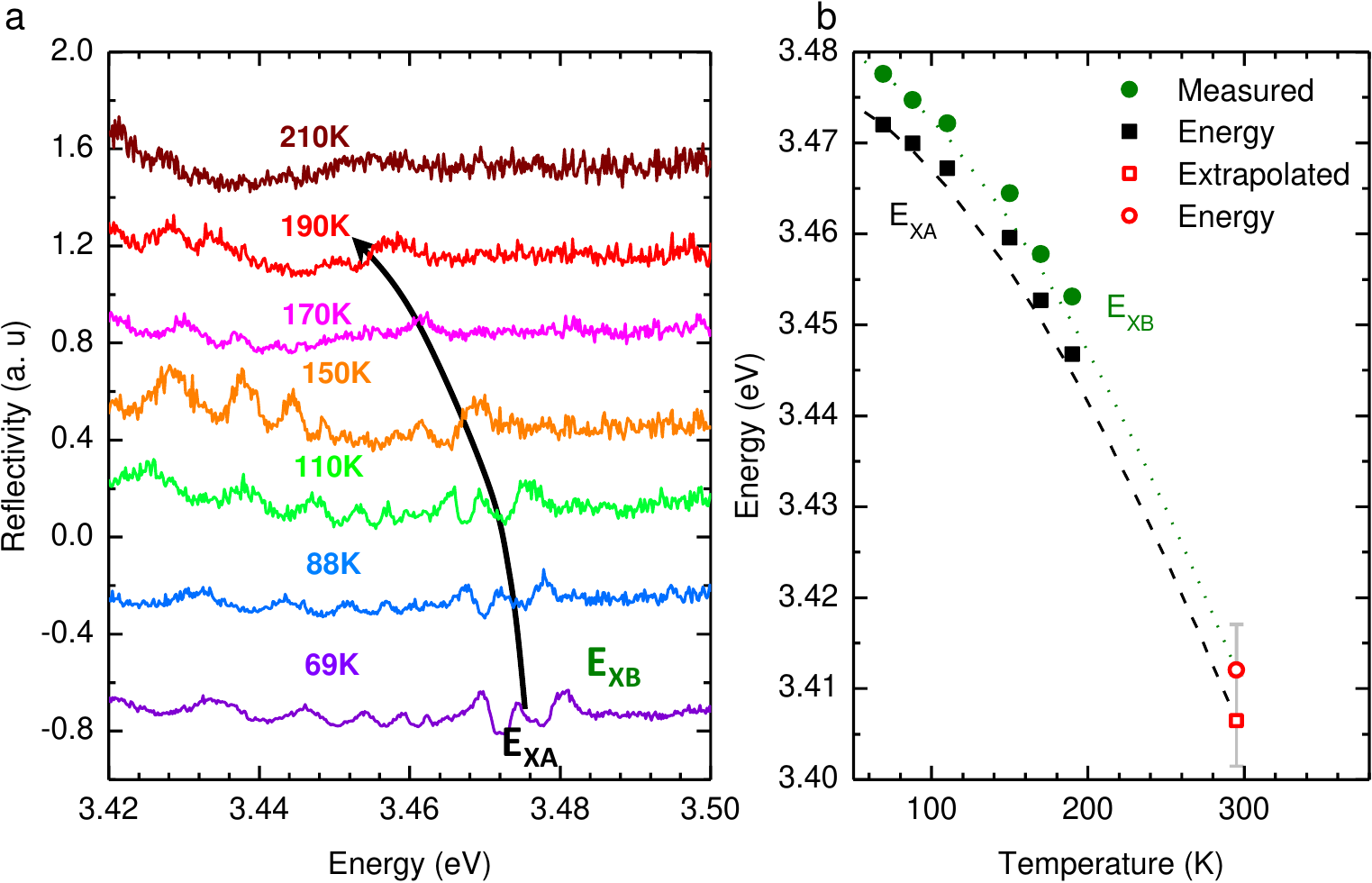}} 
  \caption{(a) Reflectivity spectrum of the GaN active layer for temperature ranging from 70 to 210K, (b) A and B excitons energies (black and green dots respectively)  as a function of the temperature, and the corresponding fit by a Varshni law (lines)} 
  \label{fig:Reflectivity}
\end{figure}

Reflectivity measurements are used to extract the exciton energy of the cavities. However, the thick p-doped AlGaN top layers (Fig.~\ref{fig:Fabrication process}a) prevent the efficient collection of the reflectivity signal coming from GaN core layer. To circumvent this limitation, an unprocessed sample is etched so that the p-AlGaN thickness is reduced from 520nm to 110nm. The Fig.~\ref{fig:Reflectivity}a shows the reflectivity spectra from 70K to 210K. The resonance of the A and B excitons are clearly identified, and reported in the panel b. From on these measurements, the exciton energies are extrapolated via a Varshni law up to room temperature (Fig.~\ref{fig:Reflectivity}b) and are equal to E$_{X_{A}}$=3.408 $\pm$ 0.005 eV and E$_{X_{B}}$=3.415 $\pm$ 0.005 eV, respectively. 

\section{Dispersion of the waveguide polaritons}\label{app:Dispersion of the waveguide polaritons}

The simulations rely on an Eigenmode Expansion (EME) method leading to the computation of the polariton and bare photon when the dielectric susceptibility of the GaN active layer includes or excludes exciton resonances \cite{Brimont_Strong_2020}. This approach allows to extract the Rabi splitting as well as the zero detuning condition $\delta_{0}$ for a given structure. The oscillator strength of the system is controlled by a tuning factor $f$, ranging between 0 and 1, accounting the screening caused by the free carriers. This is illustrated by the modelled FSR presented Fig.~\ref{fig:Emission spectrum}d, where an agreement between the experimental and modelled FSR is observed for oscillator strength of $f=60\%$ under electrical injection ($f=55\%$ under optical pumping in Fig.~\ref{fig:Optical_pumping}b, respectively).
\\
Additionally, based on these dispersions and using a quasi-particle model, the excitonic and photonic Hopfield coefficient $C$ and $X$ are also determined.

\section{Evolution of the threshold in conventional laser}

In a conventional laser, the laser condition results from the reciprocity between the absorption and the gain by the electrons and holes in the active medium, described by the Bernard-Durrafourg condition \cite{Bernard_Laser_1961}. Considering a partial pumping of a cavity at $T=0$ \cite{Souissi_Ridge_2022}, the gain at threshold writes:

\begin{equation}
    \gamma_{e-h,th}=(\alpha_{0}+\alpha_m) \frac{L_{cav}}{2L_{gain}-L_{cav}}
\end{equation}

where $\alpha_0$ is the electron-hole absorption at zero carrier density, $\alpha_m$ is the mirror loss, L$_\text{gain}$ is the length pumped region and L$_\text{cav}$ the total length of the cavity. This expression diverges for L$_\text{gain} \rightarrow 0.5$, as represented in Fig.~\ref{fig:Laser threshold}a.


%


\bibliography{Extraction}

\end{document}